\documentclass[twocolumn,aps,showpacs,preprintnumbers,amsmath,amssymb]{revtex4}
\usepackage{graphicx}
\usepackage{color}
\usepackage{dcolumn}
\usepackage{bm}
\newcommand{\mev}{\, \text{MeV}}
\newcommand{\gev}{\, \text{GeV}}

\newcommand{\bs}[1]{\mbox {\boldmath $#1$}}
\begin{document}

{\parbox[b]{1in}{\hbox{\tt INT-PUB-09-011}}}
\title{$^6$He $\beta$-decay rate and the suppression of the axial constant in nuclear matter}
\author{Sergey Vaintraub}
\email{sergey.vaintraub@mail.huji.ac.il}
\affiliation{Racah Institute of Physics, The Hebrew University, Jerusalem, 91904, Israel}
\author{Doron Gazit}
\email{doron.gazit@mail.huji.ac.il}
\affiliation{Institute for Nuclear Theory, University of Washington, 
Box 351550, 98195 Seattle, Washington, USA}
\author{Nir Barnea}
\email{nir@phys.huji.ac.il}
\affiliation{Racah Institute of Physics, The Hebrew University, Jerusalem, 91904, Israel}
\date{\today}

\begin{abstract}
We present a microscopic calculation of the $^6$He $\beta$-decay into the ground
state of $^6$Li. To this end, we use chiral
perturbation theory at next-to-next-to-next-to-leading
order to describe the nuclear weak-currents. The nuclear wave 
functions are derived from the J-matrix inverse scattering nucleon-nucleon
potential (JISP), and the Schr\"{o}dinger equation is solved using the
hyperspherical-harmonics expansion. Our calculation brings the theoretical
decay-rate within $3\%$ of the measured one. This success is attributed to the
use of chiral-perturbation-theory based mesonic currents, whose contribution is
qualitatively different compared to standard nuclear physics approach, where
the use of meson exchange currents worsens the comparison to experiment. 
%New sentence 
The inherent inconsistency in the use of the JISP potential together with chiral-perturbation-theory based is argued not to affect this conclusion, though a more detailed investigation is called for.
%New sentence ends
We conclude that any suppression of the axial constant in nuclear matter is
included in this description of the weak interaction in the nucleus. 
\end{abstract}


\maketitle
\section{Introduction}
$\beta$-decay is the every-day reflection of weak-interaction in nuclei. As
such, it provides an experimental window to the properties of the weak
interaction at nuclear density. 

In particular, theoretical studies of $\beta$-decay rates of nuclei
have argued for a suppression of the axial coupling constant $g_A$,
from its vacuum value, as extracted from the lifetime of the neutron
$g_A=1.2695 \pm 0.0029$ \cite{PDBook}, to unity, i.e. $g_A =1$
\cite{BuckPhysRevLett1983,WilkinsonNuPhA1973}. According to a recent study,
this suppression occurs gradually, as the mass of the nucleus grows, and
fully utilized for $A\approx 40$ \cite{ChouPhysRevC1993}. 

The ramifications of this suppression are numerous. For example, to the understanding
of astrophysical phenomena, such as neutron-star cooling
and core collapse supernovae, whose dynamics is controlled by the weak
interactions. It is of no surprise that the source of this suppression
has been the target of many theoretical works, which have associated
it with a partial restoration of chiral symmetry in finite densities, deficiencies in the inclusion of correlations between
nucleons, loop-corrections to the axial current originating in
nucleonic excitations and mesonic currents, or combination
of the three \cite{FranzRevModPhys1992,1994JPhGBirse,1985PhysRevLettRho, CaurierRevModPhys2005,WeiseNuPhA1993}. 

An important assumption is hidden in these suppression
mechanism: if a full calculation of the weak interaction inside
nuclei was possible from first principles, then the calculated decay-rates should
agree with the experimental ones. That is, if one could describe
correctly the correlations between nucleons, and the weak interaction
of an external probe with a nucleus, then one should recover the
physical value of the axial constant. In order to do that,
pertinent is to solve the nuclear problem from first-principles. Due to the
strong correlations involved in the problem, a calculation of nuclear
wave functions from the nucleonic degrees of freedom is at reach only
for very light nuclei. 

The lightest nucleus that undergoes a $\beta$-decay is the
triton. However, the theory cannot be checked in the triton since its
half-life is used to remove some freedom in the weak interaction of a
lepton with a nucleus, as will be explained explictly later. The
lightest nucleus that can provide a test to the theory is thus
$^6$He. $^6$He ($\rm{J}^\pi=0^{+}$) is an unstable nucleus, which undergoes
a $\beta$ decay with a half-life $\tau_{1/2}=806.7 \pm
1.5\,\text{msec}$ to the ground state of $^6$Li ($\rm{J}^\pi=1^{+})$
\cite{AjzenbergSelove:1988ec}.

However, a microscopic calculation of $^6$He from its
nucleonic degrees of freedom, failed to reproduce the $\beta$-decay
rate. This study, accomplished by Schiavilla and Wiringa
\cite{SchiavillaPhysRevC2002}, has used the realistic Argonne $v18$
(AV18) nucleon-nucleon potential, combined with the Urbana-IX (UIX)
three-nucleon-force (3NF), to derive the nuclear wave functions, through the
variational Monte-Carlo approach. The model used for the nuclear weak axial
current includes one- and two-body
operators. The two-body currents are
phenomenological, with the strength of the leading two-body term -- associated
with $\Delta$-isobar excitation of the nucleon -- adjusted to reproduce the
Gamow-Teller matrix element in tritium $\beta$-decay. The calculated
half-life of $^6$He overpredicts the measured one by about 9\%. An
unexpected result of the calculation, was that two-body currents lead to a
$1.7\%$ increase in the value of the Gamow-Teller matrix element of $^6$He,
thus worsening the comparison with experiment. The authors of this paper have
presumed that the origin of this discrepancy is either in the approximate
character of the VMC wave functions, or in the discrepancies of the nuclear
model of the weak interaction. 
%New Sentence
Pervin {\it et} al. \cite{PervinPhysRevC2007}, have used the GFMC approach to evolve the VMC wave functions ansatz. They showed that this brings the single nucleon Gamow-Teller matrix element to about 0-3\% deviation from the experimental value. However, the MEC are still expected to increase the deviation from the experiment to about 2-5\% from the experimental value, leaving this problem intact.
%New Sentence ends 

In the current {\it paper}, we argue that the origin of the discrepancy is
indeed in the model of the weak interaction inside the nucleus. The foundation
of such an argument has to be in the underlying theory, i.e. quantum
chromodynamics (QCD). Thus, we describe the weak currents within the nucleus,
using an effective theory of QCD, namely chiral perturbation theory ($\chi$PT),
applicable at low energies, relevant to $\beta$-decay processes
\cite{WeinbergPhA1979, WeinbergNuPhB1991, WeinbergPhysLettB1990,
  GasserAnPh1984}. We use the triton $\beta$-decay to calibrate the strength
of the contact interaction part of the meson-exchange currents, thus the
calculation is without any free-parameters. 
The six-body nuclear problem is solved in a fully {\it ab-initio} approach,
expanded in hyperspherical-harmonics function, from its nucleonic degrees of
freedom \cite{NovoselskyPhRvA1995,BarneaAnPh1997,BarneaPhRvA1998}. The
nuclear wave functions are derived from J-matrix inverse scattering
nucleon-nucleon potential (JISP), describing two-nucleon scattering data and
bound and resonant states of light nuclei to high accuracy
\cite{ShirokovPhysRevC2003,ShirokovPhysLettB2005,
  ShirokovPhysLettB2004}. Using this approach not only brings the calculated
$\beta$-decay rate to within $3\%$ of the measured data, but also changes
qualitatively the contribution of the two-body meson exchange currents (MEC)
compared to the work of Schiavilla and Wiringa. $\chi$PT based MEC are found
to decrease the Gamow-Teller matrix element, compared with the increase found
by Schiavilla and Wiringa. We argue that this qualitatively different behavior originates in the use of $\chi$PT based MEC, rather than the specific choice of the potential. 

\section{Theoretical Formalism}
We start with a brief reminder of $\beta$-decay process, and the formalism used in the calculation. 
The decay is a weak process, in which an unstable nucleus of charge $Z$ emits an
electron and anti-electron-neutrino, leaving a nucleus of charge $Z+1$. The
interaction is mediated through the exchange of heavy W$^{+}$ boson. As the
momentum transfer in the process is much smaller than the mass of the 
$W^{+}$ boson, the weak interaction Hamiltonian is given by
$\hat{H}_{W}=-\frac{G |{V_{ud}}|}{\sqrt{2}}\int {d^{3}x 
\hat{j}^{-}_{\mu }(\vec{x}) \hat{J}^{+\mu }(\vec{x})}$, where $G=1.166371(6) 
\times 10^{-11} \mev^{-2}$ is the Fermi coupling constant \cite{PDBook}, 
$V_{ud}=0.9738(4)$ is the CKM matrix element mixing $u$ and $d$ 
quarks involved in the process \cite{PDBook}, $\hat{j}^{-}_{\mu }(\vec{x})$ 
is the lepton charge lowering current, and $\hat{J}^{+\mu }$ 
is the nuclear charge raising current. The decay rate can be calculated using
Fermi's Golden rule, and it is proportional to the squared matrix element of
this weak Hamiltonian $\langle f \| \hat{H}_{W} \| i \rangle$, where $i$ ($f$)
is the initial (final) state. The lepton current is well-approximated as a
current of charged Dirac particles, thus results in kinematical factors to the
decay-rate. The weak nuclear current can be written as: $\hat{J}^{+\mu
}=\frac{\tau_{+}}{2} \left(\hat{J}^{V\mu}+\hat{J}^{A\mu}\right)$, where
$\tau_{+}$ is a Pauli matrix. $\hat{J}^{V\mu}$ ($\hat{J}^{A\mu}$) has a polar-
(axial-) vector symmetry. Here, we will discuss either a triton decay, or $^6$He
decay, hence the transitions are constrained by a selection rule on the
angular-momentum change in the transition: $\Delta J = 0,\,1$. Thus, a
multipole decomposition of the nuclear current is helpful. Due to the small
momentum-transfer only the lowest multipoles contribute, i.e. the $J=1$
electric multipole of axial-vector symmetry $E_1^A$, and in the case of triton
also the $J=0$ coulomb multipole of polar-vector symmetry $C_0^V$. We
explicitly checked that indeed the contribution of neglected multipoles to the
decay-rate of $^6$He can be bounded by $1\%$
\cite{SchiavillaPhysRevC2002,VaintraubMSc2008}. The leading order contribution
to the $E_1^A$ and $C_0^V$ operators are proportional to the Gamow-Teller and
Fermi operators, respectively. Thus, it is customary, when discussing the
experimental rates, to talk about the empirical Fermi and Gamow-Teller matrix
elements, instead of $E_1^A$ and $C_0^V$, using the relations
\begin{equation}
  \rm{F} \equiv \sqrt{\frac{4\pi}{{2J_i+1}}} {\langle C_0^V \rangle }\;,
\end{equation} 
and 
\begin{equation}
  \rm{GT} \equiv \sqrt{\frac{6\pi}{{2J_i+1}}} \frac{\langle E_1^A \rangle }{g_A}\;.
\end{equation}

Here,
$\langle C_0^V\rangle \equiv \langle f \| C_0^V \| i \rangle$, and similarly
for $\langle E_1^A\rangle$, $J_i$ is the total angular momentum of 
the initial nucleus, and $g_A\!=\!1.2695 \pm 0.0029$ is the axial constant~\cite{PDBook}.

As discussed by Simpson~\cite{SimpsonPhysRevC1987}, and later revisited by
Schiavilla {\em et al}.~\cite{SchiavillaPhysRevC1998,SchiavillaPhysRevC2002},
the ``comparative'' half-life is related to the ``empirical" GT and F operators thorough
\begin{equation}
 (fT_{1/2})_t = \frac{\, K/(G^2|V_{ud}|^2)}
                     { |\text{F}|^2+\,\frac{f_A}{f_V}g_A^2 |\text{GT}|^2}\,. 
\end{equation}
Here, $K\!=\!2\pi^3\ln{ 2}/m_e^5$ (such that
$K/(G^2|V_{ud}|^2)=6146.6 \pm 0.6 \,\rm{sec}$), and
$f_A/f_V\!=\!1.00529$~\cite{SchiavillaPhysRevC1998} accounts for the small
difference in the statistical rate function between vector and axial-vector
transitions. Putting the measured $^6$He comparative half-life
$(fT_{1/2})_t=812.8 \pm 3.7 \rm{sec}$ \cite{SchiavillaPhysRevC2002}, one
extracts $|\rm{GT}(^6{\rm He})|_{expt} = 2.161 \pm 0.005$. For
triton, $(fT_{1/2})_t=1129.6 \pm 3 \rm{sec}$ \cite{2005PhLBAkulov}, thus
$|\rm{GT}(^3{\rm H})|_{expt}\!=\!1.6560\pm0.0026$ \footnote{For
  triton $|\langle C_0^V\rangle|=0.99955(15)/4\pi$
  \cite{SchiavillaPhysRevC2002,VaintraubMSc2008}.}.  

In order to complete a calculation, we have to specify the detailed structure
of the weak-current, and to calculate the nuclear wave functions. These will
combine to produce the theoretical $E_1^A$, which will be compared to the
experimental ones above.  

%=============================================================================
\section{$\chi$PT weak currents in the nucleus}
\label{sec:MEC}
%=============================================================================
The main difference between the current work and previous ones, is the
physical origin of the currents.  
The last two decades of theoretical developments have provided us with an
effective theory of QCD, in the form of $\chi$PT \cite{WeinbergPhA1979,
  WeinbergNuPhB1991, WeinbergPhysLettB1990, GasserAnPh1984}.  The $\chi$PT
Lagrangian is constructed by integrating out QCD degrees of freedom of the
order of  $\Lambda_\chi\sim1 \gev$ and higher. It retains all assumed symmetry
principles, particularly the approximate chiral symmetry of the underlying
theory. This SU(2)$_A\times$SU(2)$_V$ symmetry is based on the small up- and
down-quarks masses (compared to the QCD breaking scale). The lack
of parity doublets in the QCD scale is interpreted as an indication that this symmetry is
spontaneously broken, with the pions as the Goldstone-Nambu bosons. Their
finite, albeit small, mass is due to the finite quark masses, explicitly
breaking the chiral symmetry.  

Furthermore, the chiral Lagrangian can be organized in terms of a perturbative expansion 
in positive powers of $Q/\Lambda_\chi$ where $Q$ is the generic momentum in
the nuclear process, i.e. the $\beta$-decay or the pion
mass~\cite{WeinbergPhA1979, WeinbergNuPhB1991, WeinbergPhysLettB1990}. The
Chiral symmetry dictates the operator structure of each term of the effective
Lagrangian, however not the coupling constants. A theoretical evaluation of
these coefficients, or low-energy constants (LECs), is equivalent to solving
QCD at low-energy, and it is not yet feasible to obtain them from lattice
calculations because of computational limitations. Alternatively, these
undetermined constants can be constrained by low-energy experiments.   

As the chiral symmetry is a gauging of the electro-weak interaction, the weak
currents are the N\"other currents of this symmetry. The weak axial current
adopted in this work is the N{\"{o}}ther current derived from the axial symmetry
of the chiral Lagrangian up to N$^3$LO~\cite{ParkPhysRevC2002,
  GazitPhD2007}. At leading order (LO) this current consists of the standard
single-nucleon part, which, as mentioned above, at low momentum transfer is proportional to the
Gamow-Teller (GT) operator, 
\begin{equation} \label{Eq:E1A_GT_LO}
E_1^A|_{\rm LO}\!=\!  i\,g_A(6\pi)^{-1/2}\sum_{i=1}^A \sigma_i \tau^{+}_i\,,
\end{equation}
where $\sigma_i$, $\tau_i^+$ are spin and isospin-raising operators of the $i$th nucleon. 

Corrections to the single-nucleon current appear at %the third order of the chiral expansion 
N$^2$LO in the form of relativistic terms. It is easily verified
\cite{ParkPhysRevC2002} that the single nucleon current achieved in the
$\chi$PT formalism, is identical to that achieved in the standard nuclear
physics approach (SNPA).  

At N$^3$LO, additional corrections appear in the form of axial MEC. While the relativistic corrections are negligible for the half life,
the MEC have a substantial influence on this $\beta$-decay rate. This is a
reflection of the fact that $E_1^A$ is a chirally unprotected operator
\cite{RhoPhysRevLett1991}.
The MEC, to this order, include two topologies: a one-(charged)-pion exchange, and a contact term (that represents, for example, two-pion exchange or the exchange of heavier mesons).
In configuration space the one-pion exchange part of the axial MEC is given by:
\begin{eqnarray} \label{eq:MEC_operator}
\lefteqn{
-\frac{2Mf_\pi^2}{g_A}\hat{\mathcal{A}}_{1\pi}^{i,a}(r_{ij})=\left.
{\mathcal{O}}_P^{i,a} y_{1\Lambda}^\pi(r_{ij})+ \right.}\nonumber \\ && {\left. +
\hat{c}_3({\mathcal{T}}_\oplus^{i,a}-{\mathcal{T}}_\ominus^{i,a})
m^2_\pi y_{2\Lambda}^\pi(r_{ij})+\right.}\nonumber \\ && {\left.
+\frac{\hat{c}_3}{3}({\mathcal{O}}_\oplus^{i,a}-
{\mathcal{O}}_\ominus^{i,a})m_\pi^2
y_{0\Lambda}^\pi (r_{ij})- \right.}\nonumber \\ && {
\left.-(\hat{c}_4+\frac{1}{4}) m^2_\pi
\left({\mathcal{T}}_\otimes^{i,a}y_{2\Lambda}^\pi(r_{ij})+\frac{2}{3}
{\mathcal{O}}_\otimes^{i,a}y_{0\Lambda}^\pi (r_{ij})\right) \right.}\nonumber .
\end{eqnarray}
Where $f_\pi\approx92.4 \mev$ is the pion-decay constant, $M\approx938.9 \mev$ is the mass of the nucleon, $m_\pi\approx 139.57 \mev$ is the charged-pion mass \cite{PDBook}, and the low-energy constants $\hat{c}_3=-3.66(8)$ and $\hat{c}_4=2.11(9)$ are calibrated in the $\pi$-N sector \cite{BernardNuclPhysB1995}. The operators used here are defined as,
\begin{eqnarray}
\vec{\mathcal{O}}^{a}_P &\equiv& -\frac{m_\pi}{4}
(\vec{\tau}^{(1)}\times\vec{\tau}^{(2)})^a
(\vec{P}_1\vec{\sigma}^{(2)}\cdot\hat{r}_{12}+
\vec{P}_2\vec{\sigma}^{(1)}\cdot\hat{r}_{12})
\nonumber \\ 
{\mathcal{O}}^{i,a}_\odot &\equiv&
(\vec{\tau}^{(1)}\odot\vec{\tau}^{(2)})^a
(\vec{\sigma}^{(1)}\odot\vec{\sigma}^{(2)})^i
\nonumber \\ 
{\mathcal{T}}^{i,a}_\odot &\equiv&
\left(\hat{r}^i_{12}\hat{r}^j_{12}-
\frac{\delta^{ij}}{3}\right){\mathcal{O}}^{i,a}_\odot \nonumber,
\end{eqnarray}
and $\odot=\times,+,-$. In addition, the Yukawa-like functions are:
\begin{eqnarray}
y^\pi_{0\Lambda}(r) &\equiv& \int\frac{d^3
k}{(2\pi)^3}e^{i\vec{k}\cdot\vec{r}}
S^2_\Lambda(\vec{k}^2)\frac{1}{\vec{k}^2+m_\pi^2}
\nonumber  \\
y^\pi_{1 \Lambda}(r) &\equiv& -\frac{\partial}{\partial r}
y^\pi_{0\Lambda}(r)
\nonumber \\
y^\pi_{2\Lambda }(r) &\equiv&
\frac{1}{m^2_\pi}r\frac{\partial}{\partial r}
\frac{1}{r}\frac{\partial}{\partial r} y^\pi_{0\Lambda}(r)
\nonumber.
\end{eqnarray}
$S_\Lambda$ is a cutoff function, which we take as a Gaussian. 

Apart from this, the MEC include a contact term, that has the form:
\begin{equation} 
\frac{2Mf_\pi^2}{g_A}\hat{\mathcal{A}}_{C}^{i,a}(r_{ij})=\hat{d}_r{\mathcal{O}}_\otimes^{i,a} \delta_\Lambda^{(3)}(\vec{r}_{ij}),
\end{equation}
where the "smeared" delta function is
\begin{equation} \label{Eq:contact}
\delta_\Lambda^{(3)}(\vec{r}) \equiv \int\frac{d^3
k}{(2\pi)^3}e^{i\vec{k}\cdot\vec{r}} S^2_\Lambda(\vec{k}^2).
\end{equation}
The LEC $\hat{d}_r$ is the only LEC up to N$^3$LO that cannot be calibrated in the single nucleon sector, as it originates in the contact interaction $\pi$-NN in the chiral Lagrangian. 
As a result, in order to determine $\hat{d}_r$, one has to
resort to a larger nuclear system. We will use the triton half-life as an
experimental datum to determine this LEC, Sec.~\ref{subsec:3H3He}.

%=============================================================================
\section {Nuclear Wave Functions} 
\label{sec:WaveFunctions}
%=============================================================================
The difference between
the one-body contribution to the $^6$He-$^6$Li GT matrix element and the
experimental value is of the order of few percent. A
result which on the one hand is very satisfying, but on the other hand implies
that numerical accuracy at a {\it per mil} level is required if we to regard the
$^6$He $\beta$-decay as a test of the MEC model.   
In view of this required level of convergence we use the JISP16 potential
\cite{ShirokovPhysLettB2005} to model the interaction between the nucleons. The JISP16 NN potential
utilizes the J-matrix inverse scattering technique to construct a soft nuclear
potential, formulated in the harmonic oscillator basis, that by construction
reproduces the NN phase shifts up to pion threshold and the binding energies of
the light nuclei with $A\leq 4$.    

We use the Hyperspherical-Harmonics (HH) expansion to solve the Schr\"{o}dinger
equation. 
The HH functions constitute a general basis
for expanding the wave functions of an $A$-body
system~\cite{FabreAnPh1983}. 
In the HH method, the translational invariant wave-function is written as
\begin{equation}  
 \Psi = \sum_{n[K]} C_{n[K]}R_{n}(\rho){\cal Y}_{[K]}(\Omega,s_i,t_i)
\end{equation}
where $\rho$ is the hyperradius, and  $R_{n}(\rho)$ are a complete set of
basis functions. The hyperangle,
$\Omega$, is a set of $3A-4$ angles, and ${\cal Y}_{[K]}(\Omega,s_i,t_i)$ are a
complete set of antisymmetric basis functions in the Hilbert space of
spin, isospin and hyperangles. The hyperradius $\rho$ is symmetric under
particle permutations since
$\rho^2=\frac{1}{2A}\sum_{i,j}(\bs{r}_i-\bs{r}_j)^2$. 
The functions ${\cal Y}_{[K]}(\Omega,s_i,t_i)$
are characterized by a set 
of quantum numbers $[K]$ \cite{BarneaAnPh1997,BarneaPhRvA1998} and
possess definite angular momentum, isospin, and 
parity quantum numbers.
They are
the eigenfunctions of the hyperspherical, or generalized angular momentum
operator $\hat{K}^2$, $\hat{K}^2{\cal Y}_{[K]}(\Omega,s_i,t_i)=K(K+3A-5){\cal
  Y}_{[K]}(\Omega,s_i,t_i)$.
The details of our method are explained thoroughly in Ref.~\cite{BarneaPhysRevC2006}.

%=============================================================================
\section {Results} 
\label{sec:Results}
%=============================================================================
\subsection{The triton $\beta$-decay - Calibration of $\hat{d}_r$}
\label{subsec:3H3He}
%=============================================================================

Our results for the ground state properties of the $A=3$ nuclei, $^3$H and
$^3$He , are presented
in table \ref{tabA3}.  
In the table, we present the energies, 
matter radii, and the leading order GT matrix element (see Eq.~(\ref{Eq:E1A_GT_LO})) as a function of $K_{max}$, the limiting value of the
hyperspherical angular momentum $K$ in
the HH expansion. As we are using the bare interaction our 
results are variational.

\begin{table}[h]
 \centering
 
 \caption{The JISP16 NN
   interaction $^{3}$He, $^{3}$H binding energies, rms matter radius, and the
   leading order GT matrix element as a function of $K_{max}$. \label{tabA3}} 
   \begin{tabular}{r|cc|cc|c}
        \hline  \hline 
     & \multicolumn{2}{c|}{$^{3}$H} & \multicolumn{2}{c|}{$^{3}$He} & \\
\hspace{0mm}  $K_{max}$       \hspace{0mm} &  
\hspace{1mm}  B.E.  \hspace{1mm} & 
\hspace{1mm}  radius \hspace{1mm}  & 
\hspace{1mm}  B.E.  \hspace{1mm} & 
\hspace{1mm}  radius \hspace{1mm}  &  
\hspace{1mm} $\rm{GT}|_{\text LO }$  \hspace{1mm}\\
                  \hline 
    4   \hspace{2mm} & 8.094  & 1.632  & 7.364 & 1.653 & 1.6656 \\
    6   \hspace{2mm} & 8.233  & 1.656  & 7.512 & 1.680 & 1.6620 \\
    8   \hspace{2mm} & 8.319  & 1.677  & 7.604 & 1.704 & 1.6575 \\
   10   \hspace{2mm} & 8.351  & 1.691  & 7.641 & 1.720 & 1.6547 \\
   12   \hspace{2mm} & 8.360  & 1.697  & 7.651 & 1.727 & 1.6538 \\
   14   \hspace{2mm} & 8.365  & 1.701  & 7.657 & 1.733 & 1.6530 \\
   16   \hspace{2mm} & 8.367  & 1.704  & 7.660 & 1.736 & 1.6526 \\
   18   \hspace{2mm} & 8.367  & 1.705  & 7.661 & 1.738 & 1.6524 \\

   \hline 
   \cite{ShirokovPhysLettB2005}$_V$ \; 
                     & 8.354 &    & 7.648       &       \\ 
   \cite{ShirokovPhysLettB2005}$_E$ \; 
                     & 8.496(20) &    & 7.797(17) &       \\ 
   \hline 
  Exp. & 8.482 &   & 7.718 &  &  \\
                \hline 
                \hline
                \end{tabular}
\end{table}

From the table it is evident that an excellent convergence is achieved for
the $A=3$
nuclei. Our results indicate that the JISP16
potential leads to an 
underbinding of about $80$keV for the $^3$He and $120$keV for the triton. Comparing our results
with the NCSM results of Shirokov {\it et} al. \cite{ShirokovPhysLettB2005},
we see a nice agreement
with their variational results \cite{ShirokovPhysLettB2005}$_V$ but a
discrepency of about $130$keV
with their effective interaction results  \cite{ShirokovPhysLettB2005}$_E$.
It should be noted that the GT matrix element converges much faster
then the matter radius. This property can be probably attributed to the fact that the GT is a medium-range operator, which is influenced by the asymptotic behavior of the wave function, described correctly using the hyperspherical functions. Comparing the JISP16 leading order GT matrix element 
with those of other potential models, see table~\ref{tab:A3GT}, we observe that
the JISP16 potential model leads to an enhancement of the 1-body matrix element
and it almost coincides with the experimental value. 
%NEW SENTENCE
This property is found also for the UCOM potential, and might be a result of the minimization of the contribution of 3NF to the binding energy, which is in the essence of both these potentials. In general, one observes from the table that non-local potentials, such as Bonn or the N$^3$LO potentials, tend to predict a value for the GT matrix element which is closer to experiment than the local potentials. 
%NEW SENTENCE ends

\begin{table}[h]
  \centering
  \caption{The dependence of the triton
    $\beta$-decay leading order GT matrix-element on the potential model.} 
  \label{tab:A3GT}
  \begin{tabular}{cc}
    \hline \hline
    Potential model &  $ \rm{GT}|_{\text LO}$  \\ \hline 
    AV18+3NF \cite{WiringaPhysRevC1995}       &  1.598(2)    \\
    Bonn+3NF \cite{MachleidtPhysRevC2001}     &  1.621(2)     \\
    Nijm+3NF \cite{StoksPhysRevC1994}        &  1.605(2)          \\
    N$^3$LO+3NF \cite{GazitUnpublished2008}        &  1.622(2)          \\
    UCOM \cite{Bacca2009}        &  1.65(1)          \\
    JISP16 [This work]                   &  1.6524(2) \\ \hline
    Expt.     & 1.656(3)           \\
                \hline \hline
   \end{tabular}
\end{table}

As explained in Sec.~\ref{sec:MEC}, we use the triton half-life as an
experimental input to determine the LEC $\hat{d}_r$. That is, we use the trinuclei
wave functions to evaluate the matrix-element $\!|\langle^3{\rm
  He}||E_1^A||^3{\rm H}\rangle|$,
of the $E_1^A$ operator built
from the $\chi$PT based weak-current,
as a function of $\hat{d}_r$, for various high-energy-cutoff values. Using the
experimentally derived value for this matrix element we get the following
calibration for $\hat{d}_r(\Lambda_\chi)$: 
\begin{eqnarray} \label{eq:d_r}
\nonumber
\hat{d}_r(\Lambda_\chi=500\mev) & = & 0.583(27)_t  (38)_{g_A} \\
\hat{d}_r(\Lambda_\chi=600\mev) & = & 0.625(25)_t  (35)_{g_A} \\ \nonumber
\hat{d}_r(\Lambda_\chi=800\mev) & = & 0.673(23)_t  (33)_{g_A}.
\end{eqnarray}  
The numbers in parenthesis denote uncertainties in the last digits. The first
error is due to the uncertainty in the triton half-life, whereas the second
one is due to uncertainty in $g_A$ (the numerical error is negligible).  

%=============================================================================
\subsection{The $^6$He-$^6$Li Gamow-Teller matrix-element}
\label{subsec:6He6Li}
%=============================================================================

Turning now to the $A=6$ case, we present in table \ref{tab:A6}
our results for the ground state properties of the $^6$He, and $^6$Li nuclei.
As evident in the table, at the value $K_{max}=14$, which corresponds to
about $2-3\cdot 10^6$ basis states, the binding energies of the 6-body nuclei
are obtained with an accuracy of few hundreds keV.
 
\begin{table}[h]
 \centering
 \caption{The JISP16 NN
   interaction $^{6}$He, $^{6}$Li binding energies, rms matter radii, and
   the leading order GT matrix element as a function of $K_{max}$. \label{tab:A6}} 
   \begin{tabular}{r|cc|cc|c}
        \hline  \hline 
     & \multicolumn{2}{c|}{$^{6}$He} & \multicolumn{2}{c|}{$^{6}$Li} & \\
\hspace{0mm}  $K_{max}$       \hspace{0mm} &  
\hspace{1mm}  B.E.  \hspace{1mm} & 
\hspace{1mm}  radius \hspace{1mm}  & 
\hspace{1mm}  B.E.  \hspace{1mm} & 
\hspace{1mm}  radius \hspace{1mm}  &  
\hspace{1mm} $ \rm{GT}|_{LO}$  \hspace{1mm}\\
                  \hline 
    4   \hspace{2mm} & 18.367  & 1.840   & 19.392 & 1.859 & 2.263 \\
    6   \hspace{2mm} & 24.103  & 1.902   & 26.124 & 1.909 & 2.247 \\
    8   \hspace{2mm} & 26.392  & 1.979   & 28.854 & 1.984 & 2.234 \\
   10   \hspace{2mm} & 27.560  & 2.051   & 30.156 & 2.051 & 2.232 \\
   12   \hspace{2mm} & 28.112  & 2.112   & 30.797 & 2.110 & 2.229 \\
   14   \hspace{2mm} & 28.424  & 2.165   & 31.132 & 2.160 & 2.227 \\
$\infty$\hspace{3mm} & 28.70(13) &       & 31.46(5) &     & 2.225(2) \\
   \hline 
   \cite{ShirokovPhysLettB2005} \; 
                     & 28.32(28) &    & 31.00(31)       &       \\ 
   \hline 
  Exp. & 29.269 & 2.18  & 31.995 & 2.09  & 2.170 \\
                \hline 
                \hline
                \end{tabular}
\end{table}

\begin{figure*}
\begin{center}
\includegraphics[scale=0.75,clip=]{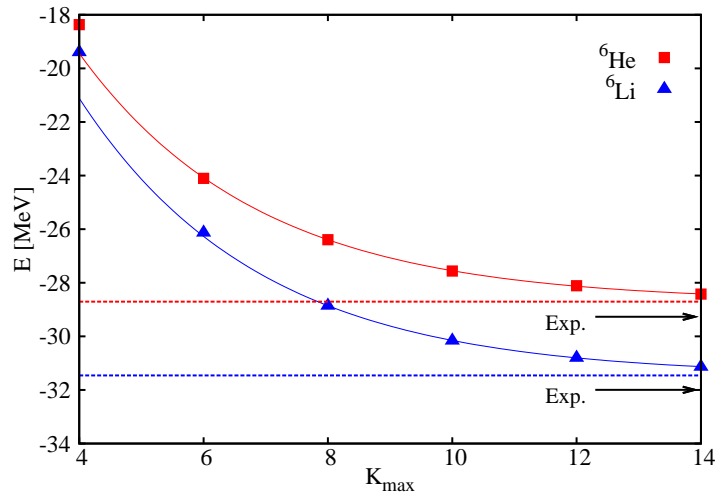}
\caption{(Color online) The convergence of the binding energies
of the $6$-body nuclei, $^6$Li and $^6$He. The continuous lines
are the fits $E(K_{max})=E_{\infty}+A e^{-\alpha K_{max}}$. The dashed lines
are the extrapolated values $E_{\infty}$. The experimental values are marked with black arrows.}
\label{fig:6BodyBE}
\end{center}
\end{figure*}

\begin{figure*}
\begin{center}
\includegraphics[scale=0.75,clip=]{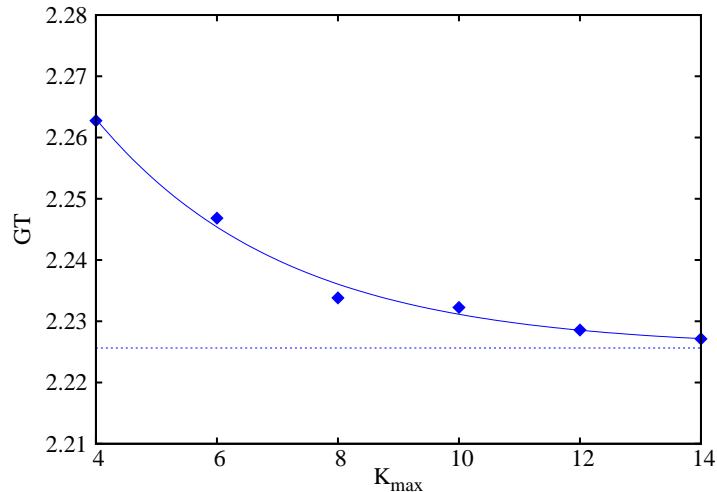}
\caption{(Color online) The convergence of the GT matrix element
for the  $^6$He-$^6$Li $\beta$-decay. The continuous line
is the fit $\rm{GT}(K_{max})=\rm{GT}_{\infty}+Be^{-\beta K_{max}}$,
the dashed line is the extrapolated value $\rm{GT}_{\infty}$.}
\label{fig:6BodyGT}
\end{center}
\end{figure*}

Taking a closer look at the table, we find that
the binding energies exhibit an exponential
convergence. Deploying this observation we 
extrapolate our results to the $K_{max}\longrightarrow \infty$ limit, using
the formula $E(K_{max})=E_{\infty}+A e^{-\alpha K_{max}}$. 
Fitting the parameters $E_{\infty},A,\alpha$ to the entries of
table~\ref{tab:A6} in the range 
$K_{max}\geq 6,8,10$ we find a rather stable value for $E_{\infty}$ with
variance of about $50$keV for $^6$Li and $130$keV for $^6$He.
The resulting binding energies are $28.70$MeV for $^6$He and
$31.46$MeV for $^6$Li. While these results are roughly $550$keV below the
experimental values, the difference $\Delta E = 2.76$MeV between the binding
energies of the two nuclei differs by merely $34$keV from the experimental
value, $\Delta E = 2.726$MeV.
In the last column of  
table~\ref{tab:A6} we present our $^6$He-$^6$Li leading order GT transition
matrix element, i.e. at the 1-body level. It can be seen 
that
the convergence pattern of the matrix element is not regular. Extrapolating
its value using the expression
$\rm{GT}(K_{max})=\rm{GT}_{\infty}+Be^{-\beta K_{max}}$
for $K_{max}\geq 0$, we get $\rm{GT}_{\infty}=2.225(2)$.
The fits of the extrapolation formulae to the calculated values are presented
in Fig.~\ref{fig:6BodyBE} 
for the binding energies and in Fig.~\ref{fig:6BodyGT} for the GT matrix-element. 

The value $\rm{GT}=2.225(2)$ we obtained for the JISP16 potential,
is in accordance with the values $\rm{GT}=2.28$ for AV8'/TM'(99) and
$\rm{GT}=2.30$ for AV8' obtained by Navratil and Ormand \cite{NavratilPhysRevC2003},
$\rm{GT}=2.28$ for the N$^3$LO NN-force of Navratil and Caurier
\cite{NavratilPhysRevC2004}, $\rm{GT}=2.25$ for AV18/UIX of Schiavilla and Wiringa
\cite{SchiavillaPhysRevC2002}, and $\rm{GT}=2.16-2.21$ for AV18/IL2 by
Pervin {\it et} al. \cite{PervinPhysRevC2007}. Moreover, it can be seen that our
accuracy in estimating the GT matrix element is at the level of {\it per mil}.
Such an accuracy enables us to disentangle numerics from physics and validates the use of
the $^{6}$He $\beta$-decay as a testing ground for an axial MEC model. 

Incorporating the $\chi$PT based contributions to the weak-current we can
finally calculate the full $^6$He-$^6$Li GT matrix-element at the N$^3$LO level. In
table~\ref{tab:A6GTfull}, we present the transition matrix-elements as a
function of $K_{max}$ and the cutoff $\Lambda_\chi$. The appropriate values of
$\hat{d}_r$ are taken from Eq. (\ref{eq:d_r}). 
Two important observations can be drawn from the table, (i) the numerical accuracy of
the calculated GT matrix-element is few per mil, and (ii) there is only a very
weak dependence on the cutoff $\Lambda_\chi$, which is of the same order of
magnitude. The second observation implies that there is no
need to refine our calculation, and moreover, the contribution of higher order
$\chi$PT corrections to the weak-current are negligible.

\begin{table}[h]
  \centering
  \caption{The dependence of the full (1-body+2-body) $^6$He-$^6$Li GT matrix-element 
    on $K_{max}$ as a function of the cutoff $\Lambda_\chi$, at the N$^3$LO level.} 
  \label{tab:A6GTfull}
  \begin{tabular}{cccc}
    \hline \hline
\hspace{0mm}  $K_{max}$       \hspace{0mm} &  
\hspace{0mm}  $\Lambda_\chi=500 \rm{MeV}$ \hspace{0mm} &
\hspace{0mm}  $\Lambda_\chi=600 \rm{MeV}$ \hspace{0mm} &
\hspace{0mm}  $\Lambda_\chi=800 \rm{MeV}$ \hspace{0mm} \\
\hline
   4  &  2.1870  &  2.1798  &  2.1703  \\
   6  &  2.1850  &  2.1805  &  2.1746  \\
   8  &  2.1868  &  2.1850  &  2.1826  \\
  10  &  2.1937  &  2.1932  &  2.1927  \\
  12  &  2.1951  &  2.1952  &  2.1955  \\
  14  &  2.1970  &  2.1975  &  2.1983  \\
\hline \hline
 \end{tabular}
\end{table}

Summarizing, the predicted GT of $^6$He is:
\begin{equation}
|\rm{GT}(^6{\rm He})|_{theo}=2.198 (1)_\Lambda (2)_N (4)_t
(5)_{g_A} = 2.198 \pm 0.007
\end{equation}
The first error is the cutoff variation dependence, the second is numerical,
the third is due to uncertainties in the triton half-life, and the last is due
to uncertainties in $g_A$. This should be compared to the experimental
matrix-element $|\rm{GT}(^6{\rm He})|_{expt}=2.161\pm 0.005$. Thus, the theory overpredicts
GT by about $1.7\%$.

\section{Discussion}

The use of phenomenologically based potential, JISP, combined with a $\chi$PT
based MEC, is an inconsistency inherent to our calculation. Clearly, the
chiral Lagrangian can be used to derive the nuclear forces as well. This
inconsistency, however, allows us to overcome limited computational resources, 
as well as theoretical difficulties (the N$^3$LO nuclear potential has not been 
fully developed yet), and to accomplish the task of a microscopic
calculation of a six-body problem. The use of a hybrid approach, sometimes
coined EFT$^{*}$, has had great success in the literature
\cite{ParkPhysRevC2002, GazitPhD2007,GazitPhLtB2008}. In all these checks, the
phenomenological nuclear forces included realistic potentials, the AV18+UIX
force model. This potential, though different than the $\chi$PT force models
in the short-range character of the force, has the correct long-range
behavior, due to the pion exchange. The JISP potential, however, is different
in this respect, as it is built in an {\it ab-exitu} approach, and does not
have an asymptotic long pion behavior, thus not consistent with chiral
symmetry even at long-distances. In addition, the JISP potential does not include
a three-body force. 

It is hard to estimate the effect of these approximations. However, in a
recent work \cite{GazitUnpublished2008}, the triton $\beta$-decay process was
calculated using force model and current derived consistently from the same
$\chi$PT N$^3$LO Lagrangian. One of the conclusions of this work has been that
the short-range correlations of the force and the short-range correlations of
the weak current are not correlated, thus the effect of the three-body-force
is negligible for GT-type operators. In addition, the JISP potential
successfully reproduces nucleon-nucleon scattering data, and the binding
energies of $A<16$ mass nuclei. However, the most convincing reason to believe
the stability of the current results, is the minimal dependence of the
half-life in the cutoff.  

We thus believe that even in the current calculation, the effect of the
approximation will not change qualitatively the results, and the effect of the
MEC. The qualitative difference originates in the different structure of the SNPA and $\chi$PT based MEC. 

A careful analysis of the difference between the MEC originating in $\chi$PT and those used in SNPA, has been accomplished by Park {\it et al} \cite{ParkPhysRevC2002}. They have shown that one-pion exchange term exists in both models. Of particular importance is the part of this term in the SNPA based MEC that represents the exchange of a pion due to a delta excitation of the nucleon, which is found to correspond roughly to the $\hat{c}_3$ term in the $\chi$PT based MEC. The coupling constant of this term $g_{\pi N \Delta}$ has been fixed by Schiavilla and Wiringa \cite{SchiavillaPhysRevC2002}, so that the theory would reproduce the triton half-life.

However, differences between the approaches arise in their short range
character. In the SNPA approach, these correspond to the exchange of a
$\rho$-meson. Such a term does not exist in the $\chi$PT approach as it arises
only at N$^5$LO \cite{ParkPhysRevC2002}. Moreover, a contact interaction of
the form of Eq.~(\ref{Eq:contact}) does not appear in the SNPA approach. It is
this contact interaction that creates the qualitative difference between the
current work and that of Schiavilla and Wiringa \cite{SchiavillaPhysRevC2002}.

\begin{figure}[h]
\begin{center}
\includegraphics[scale=0.65,clip=]{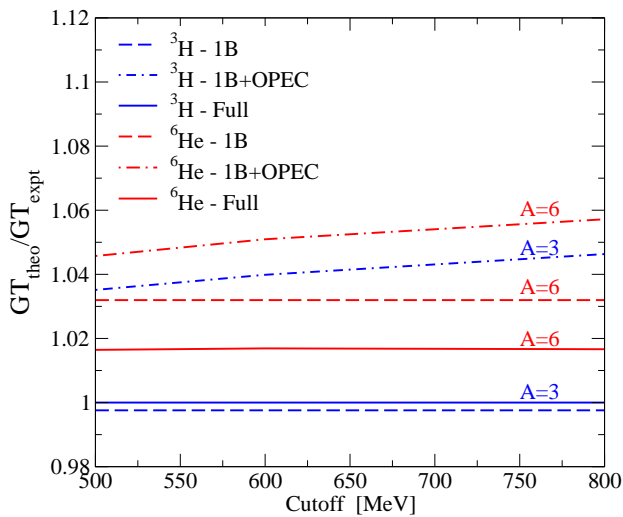}
\caption{(Color online) Relative contributions to the theoretical GT matrix
  elements as a
  function of the EFT cutoff. All the results are normalized to the empirical
  values. The blue lines indicated by $A=3$ correspond to the $^3$H-$^3$He
  $\beta$-decay. The red lines indicated by $A=6$ correspond to the
  $^6$He-$^6$Li case. Dashed lines correspond to 
  the 1-body impulse approximation (1B). Dashed-dotted lines correspond
  to 1-body plus one-pion exchange current (OPEC). Continuous lines correspond to
  full calculation (note that in the case of $^3$H this is calibrated to give
  exactly the experimental value). } 
\label{fig:relative_importance}
\end{center}
\end{figure}

In order to acknowledge that, we plot in Fig.~\ref{fig:relative_importance}
the relative contribution of each the terms, i.e. one-body, one-pion-exchange
and full calculation, to the GT matrix element. One first recognizes that the
one-pion contribution to the matrix element has a positive sign in both $^3$H
and $^6$He, and that the contact interaction has a negative contribution to the matrix element. In the case of $^3$H this is only a partial cancellation, as it is
calibrated to increase the 1-body matrix element and to bring the calculation
into the experimental value. 
%New sentence
In view of the fact that the one-body calculation in the case of the JISP potential almost exhausts the total GT strength, one might suggest that the negative sign of the contact interaction, as well as the partial cancellation is an artifact of the potential. However, the same partial cancellation is found also in a consistent N$^3$LO calculation of $^3$H decay, thus it is not a result of the use of the JISP potential \cite{GazitUnpublished2008}. 

% New Sentence ends. Begin change of sentence
In contrast to $^3$H, when examining the case of $^6$He, one observes that the 
% end of changed sentence - return to old version.
negative contribution of the contact
term is bigger (in absolute value) than the one-pion-exchange contribution,
thus leading to a total negative contribution of the MEC. This negative
contribution is needed as the single-nucleon GT is bigger than the
experimental GT.

Recalling the fact that the SNPA approach does not contain a contact interaction, we understand the origin of the positive contribution of the MEC in that approach, which increases the difference between the calculated and measured decay-rates. 

\section{Summary and Conclusions}

In this work we have used the $^6$He $\beta$-decay as a testing ground
for the nuclear weak-current derived from $\chi$PT. A precondition for such a
task is an accurate evaluation of the $^6$He-$^6$Li weak transition
matrix-element at the per mil level. To this end we have used
the soft NN potential JISP16 to describe the nuclear dynamics and
the HH expansion method to solve the Schr\"{o}dinger equation. The weak
interaction in the nucleus is completely determined by fixing the short range
behavior of the scattering operator to reproduce the experimental $^3$H
half-life, resulting in a parameter-free prediction of the $^6$He $\beta$-decay
rate. 

We have found that at the 1-body, impulse approximation, level
the $^{6}$He-$^{6}$Li GT matrix-element is over predicted by roughly $3\%$. 
This observation for the JISP16 potential is in agreement with previous findings
for other potential models.
Adding 2-body, meson-exchange, currents derived within $\chi$PT, we have found that
in contrast with the previous work of Schiavilla and Wiringa \cite{SchiavillaPhysRevC2002}, the 2-body MEC
contribution to the $^{6}$He-$^{6}$Li transition matrix element is
negative. We argue that this difference originates in the different
short-range character of the MEC derived in the two approaches. We find that
both for $^3$H and $^6$He, there is a sign difference between the positive
contribution of the long-range one-pion-exchange current, and the negative
contribution of the contact interaction in $\chi$PT, representing higher
degrees of freedom which were integrated out in the development of the
effective theory. In the case of the 6-body transition, however, the contact
interaction has a bigger value than the one-pion exchange contribution. Thus,
it provides the origin to the sign difference between the MEC contribution in
$^3$H and $^6$He. This contact interaction does not exist in the standard
nuclear physics approach, adopted by Schiavilla and Wiringa. Therefore, the
reconciliation between the theoretical and the experimental $^{6}$He half-life
is due to the use of the $\chi$PT formalism. 

Our calculation points to an agreement at the level of about 1.7\% between the
measured and calculated GT matrix elements. This result should be contrasted
with the difference of 5.4\% obtained by Schiavilla {\it et}
al. \cite{SchiavillaPhysRevC2002}. More importantly, it shows that dominant
contributions that arise naturally in the $\chi$PT formalism, and do not
appear in the standard nuclear physics approach, are essential to a successful
prediction of this weak observable. In order to pin-point this argument, a use
of a consistent approach, in which both the weak currents and the nuclear
forces are derived from the same microscopic theory, is called for. 

The agreement between the calculated and measured decay rates of
$^6$He indicates that there is no signature in this observable for an
additional suppression of the axial constant. It appears that all the needed
suppression originates in correlations between nucleons in the nucleus,
revealing itself in the form of exchange currents.  

\section*{Acknowledgments}
We thank Rocco Schiavilla for helpful discussions. 
The work of S.V.\ and N.B.\ was supported by the ISRAEL SCIENCE FOUNDATION 
(Grant No.~361/05).
D.G.\ acknowledges support from U.\ S.\ DOE Grant DE-FG02-00ER41132. 

%%%%%%%%%%%%%%%%%%%%%%%%%%%%%%%%%%%%%%%%%%%%%%%%%%%%%%%%%%%%%%%%%%%%%%%%

\end{document}